\documentstyle[12pt]{article}

\begin{document}
\title{Horizon Instability in the
Cross Polarized \\Bell - Szekeres Spacetime}
\author{ Ozay Gurtug and Mustafa Halilsoy\\
 Department of Physics, Eastern Mediterranean University\\
 G. Magusa, North Cyprus, Mersin 10 - Turkey\\
 email: ozay.as@mozart.emu.edu.tr\\
\date{}\\}
\maketitle
\begin{abstract}
{\small The quasiregular singularities (horizons) that form in the collision
of cross polarized electromagnetic waves are, as in the linear polarized
case, unstable. The validity of the Helliwell-Konkowski stability conjecture
 is tested
for a number of exact backreaction cases. In the test electromagnetic case
the conjecture fails to predict the correct nature of the singularity
while in the scalar field and in the null dust cases the aggrement is justified.}
\end{abstract}
\newpage

\section{Introduction}

It has been known for a long time that owing to planar property and mutual
focussing, colliding plane waves (CPW) result in spacelike singularities [1]. These
singularities are somewhat weakened when the waves are endowed with a
 relative
cross polarization prior to the collision. A solution given by Chandrasekhar
 and
Xanthopoulos (CX) [2], however constitutes an example contrasting this category, namely,
it possesses a Cauchy Horizon (CH) instead of a spacelike singularity.
Naturally, this solution initiated a literature devoted entirely on the quest
of stability of horizons formed hitherto. CH formed in spacetimes of CPW was
 shown
by Yurtsever to be unstable against plane-symmetric perturbations [3].
 A linear
perturbation analysis by CX reveals also an analogues result [4]. Any such
perturbation applied to a CPW spacetime will turn the CH into an essential
singularity.

A second factor that proved effective in weakening the strength of a singularity
in CPW is the electromagnetic (em) field itself. In other words, the degree
 of divergence
in the curvature scalars of colliding pure gravitational waves turn out to be
stronger than the case when em field is coupled to gravity. In particular,
collision of pure em waves must have a special significance as far as
singularity formation is concerned. Such an interesting solution was given
by Bell and Szekeres (BS) which describes the collision of two linearly polarized
step em waves [5]. The singularity (in fact a CH) formed in the interaction
region of the BS solution was shown to be removable by a coordinate transformation.
 On the null boundaries, however it possesses esential curvature singularities which
 can not be removed by any means. Since cross polarization and em field both
 play roles in the nature of resulting singularity it is worthwhile to purse
 these features together. This invokes a cross polarized version of the BS
 (CPBS) solution which was found long time ago [6,7]. This metric had the nice feature
 that the Weyl scalars are all regular in the interaction region. Cross polarization,
  however,
 does not remove the singularities formed on the null boundaries. In
 this paper we choose CPBS solution as a test ground, instead of BS, with
 various added test fields to justify the validity of a CH stability conjecture
 proposed previously by Helliwell and Konkowski (HK) [8,9]. Unlike the tedious
 perturbation analysis of both CX and Yurtsever the conjecture seems to be
 much economical in reaching a direct conclusion about the stability of a CH.
  This is our main motivation for considering the problem anew, for the
 case of untested solutions in CPW. In this paper we look at the spacetimes: a) single
 plane wave with added colliding test fields and b) colliding plane waves
  having
 non-singular interaction regions with test field added, Fig.1 illustrates these
 cases.

 The terminology of singularities should be follwed from the classification
 presented by Ellis and Schmidt [10]. Singularities in maximal four dimensional
 spacetimes can be divided into three basic types: quasiregular (QR),
 scalar curvature (SC) and non-scalar curvature (NSC). The CH stability conjecture
 due to HK is defined as follows.
 \newtheorem{guess}{Conjecture}
 \begin{guess}
 For all maximally extended spacetimes with CH, the backreaction due to a field
 (whose test stress-energy tensor is $T_{\mu \nu}$) will affect the horizon
 in one of the following manners.
 a)If $T_{\mu}^{\mu}$, $T_{\mu \nu}T^{\mu \nu}$ and any null dust density
 $\rho$ are finite, and if the stress energy tensor $T_{ab}$ in all parallel
 propagated orthonormal (PPON) frames is finite, then the CH remains non-singular.
 b)If $T_{\mu}^{\mu}$, $T_{\mu \nu}T^{\mu \nu}$ and any null dust density
 $\rho$ are finite, but $T_{ab}$ diverges in some PPON frames, then an
  NSC singularity will be formed
 at the CH.
 c)If $T_{\mu}^{\mu}$, $T_{\mu \nu}T^{\mu \nu}$ and any null dust density
 $\rho$ diverges, then an SC singularity will be formed at the CH.
 \end{guess}
 Expressed otherwise, the conjecture suggests to put a test field into the
  background
 geometry and study the reaction it will experience. If certain scalars
  diverge then
 in an exact back-reaction solution the field will respond with an infinite
  strength
 to the geometry (i.e action versus reaction). Such an infinite back-reaction
  will
 render a CH unstable and convert it into a scalar singularity. \\

The paper is organized as follows. In section II, we review the CPBS solution
 and the
correct nature of the singularity structure is presented in Appendix A.
 Section III,
deals with geodesics and test em and scalar field analyses. In section IV,
 we present an exact
back reaction calculation for the collision of cross polarized em field
 coupled with scalar field. The
derivation of Weyl and Maxwell scalars are given in Appendix B. The insertion of
test null dusts to the background CPBS spacetime and its exact back reaction
 solution is studied
in section V. Appendix C is devoted for the properties of this solution.
 The paper is concluded with a
discussion in section VI.

\section{The Cross-Polarized BS (CPBS) Metric}

The metric that describes collision of em waves with the cross polarization
was found to be [7] \\
\begin{equation}
ds^{2}=F\left(\frac{d\tau^{2}}{\Delta}-\frac{d\sigma^{2}}{\delta}\right)-\Delta
 F dy^{2}-\frac{\delta}{\Delta}(dx-q\tau dy)^{2}
\end{equation}

In this representation of the metric our notations are\\
\begin{eqnarray}
\tau &=& \sin(au + bv) \nonumber\\
\sigma &=& \sin(au - bv)\nonumber\\
\Delta &=& 1- \tau^{2}\nonumber\\
\delta &=& 1- \sigma^{2}\nonumber\\
2F &=& \sqrt{1+q^{2}}(1+ \sigma^{2}) + 1 - \sigma^{2}
\end{eqnarray}
in which $0 \leq q \leq 1$ is a constant measuring the second polarization,
 $(a,b)$
are constant of energy and $(u,v)$ stand for the usual null coordinates.
It can be seen easily that for $q=0$ the metric reduces to BS. Unlike
the BS metric ,however, this is conformally non-flat for $( u>0, v>0)$, where
the conformal curvature is generated by the cross polarization. As a matter
of fact this solution is a minimal extension of the BS metric. A completely
 different
generalization of the BS solution with second polarization was given by
 CX [11]. Their
solution, however, employs an Ehlers transformation and involves two essential
parameters which is therefore different from ours. Both solutions form
 CH in the
interaction region. Our result drown out in this paper, namely, that the
horizon is unstable against added sources can also be shown to apply to the
CX metric as well. As it was shown before the interaction region
 $(i.e\, u>0,v>0)$
of the above metric is of type-D without scalar curvature singularities. We
wish to check now the possible singularities of metric (1). The single
 component
of the Weyl scalar suffices to serve our purpose. We find that the real part
of the Weyl scalar $\Psi_{4}$ is given by
\begin{eqnarray}
Re \Psi_{4} &=& -\left(\frac{a \delta(u) \tan(bv)}{2( \Sigma^{2}
+q^{2} \sin^{2}(bv))}\right)\left[\cos^{4}(\alpha/2)
+\frac{q^{2}}{4}(5 \sin^{2}bv-3)\right. \nonumber \\
& & \nonumber \\
& &\left.- \cos(2bv) \sin^{2}(bv) \sin^{4}(\alpha/2)\right]
\end{eqnarray}
where we have used the abbreviations
\begin{eqnarray}
\Sigma&=&\cos^{2}(\alpha/2) + \sin^{2}(bv) \sin^{2}(\alpha/2) \nonumber \\
q&=&\sin \alpha
\end{eqnarray}
As $q\rightarrow 0\, (or \alpha \rightarrow 0)$ we obtain
\begin{equation}
Re \Psi_{4} = \Psi_{4}=- \frac{a}{2} \delta(u) \tan(bv)
\end{equation}
which reduces to the singularity form of the BS spacetime given by
 $ u=0,bv=\pi/2$.
We see that the same singularity remains unaffected by the
 introduction of the cross
polarization. A similar calculation for $ Re \Psi_{0} $ gives the symmetrical
 singular hypersurface sitting on $ v=0, au= \pi/2 $. Now in order to explore
 the true nature of the singularity we concentrate our account on the incoming
 region II $ (u>0, v<0)$. The metric in this region is expressed in the form
 \begin{equation}
 ds^{2}=2e^{-M}dudv-e^{-U}[(e^{V}dx^{2} + e^{-V}dy^{2})\cosh W -2\sinh W\,dxdy]
 \end{equation}
 where
 \begin{eqnarray*}
 e^{-M}&=&2F=1+\sqrt{1+q^{2}}+\left(\sqrt{1+q^{2}}-1\right)\sin^{2}(au) \\
 e^{V}\cosh W&=&\frac{1}{F} \\
 e^{-V}\cosh W&=&F+\frac{q^{2}\sin^{2}(au)}{F}
 \end{eqnarray*}
We observe that for $q\neq 0$, $F(u)$ is a bounded positive definite function
which suggests that no additional singularities arise except the one occuring
already in the BS case, namely at $ au=\pi/2, v=0 $. To justify this we have
 calculated
all Riemann components in local and PPON frames (see Appendix A ). It is
 found that
all Riemann tensor components vanish as $ au \rightarrow \pi/2 $. In the
PPON frame, however, they are all finite and according to the classification
scheme of Ellis and Schmidt such a singularity is called a quasiregular
(QR) singularity. This is said to be the mildest type among all types of
 singularities.
To check whether the QR is stable or not we consider generic test fields
 added
to such a background geometry and study the effects. This we will do in the
 follwing sections.

\section{Geodesics Behaviour, Test em and Test Scalar Fields}
We are interested in the stability of QR singularities that are
developed at $ au=\pi/2$ in region II and $ bv=\pi/2$ in region III. To
 investigate their stability
we will express geodesics and behaviour of test em and scalar fields by
 calculating
stress - energy tensor in local and PPON frames. \\
Our discussion on geodesics will be restricted in Region II only.
 We shall consider the
geodesics that originate at the wave front and move toward the quasiregular
singularity. Solution of geodesics equations in region II can be obtained by
 geodesics Lagrangian
method and using $u$ as a parameter. The results are
\begin{eqnarray}
x-x_{0}&=&-\frac{2P_{x_{0}}\left[ 1+2q^{2}\right]}{a}\tan(au) +
 \frac{3P_{x_{0}}\left[5q^{2}+2-2\sqrt{1+q^{2}}\right]u}{4} \nonumber \\
& & - \frac{P_{x_{0}}\left[\sqrt{1+q^{2}}-1\right]^{2}}{8a} \sin(2au) -
 \frac{2P_{y_{0}}q}{a\cos(au)} \nonumber \\
y-y_{0}&=&-\frac{2qP_{x_{0}}}{a\cos(au)}-\frac{2P_{y_{0}}\tan(au)}{a}
 \nonumber \\
v-v_{0}&=&\frac{\tan(au)}{a}\left[P^{2}_{x_0}(1+2q^2) +P^{2}_{y_{0}}\right]
 +u\left[\frac{\epsilon}{4}(1+3\sqrt{1+q^{2}})\right. \nonumber \\
& &\left.-\frac{3P^2_{x_0}}{8}(5q^{2}+2-2\sqrt{1+q^{2}})\right] +
 \frac{2P_{x_{0}}P_{y_{0}}q}{a\cos(au)}   \nonumber \\
& & + \frac{( \sqrt{1+q^{2}}-1) \sin(2au)}{8a}\left[\frac{P^{2}_{x_{0}}}{2}(
 \sqrt{1+q^{2}}-1)-\epsilon\right]
\end{eqnarray}
where $\epsilon=0$ for null and $ \epsilon=1$ for time like geodesics and
$x_{0},y_{0},v_{0},P_{x_{0}}$ and $P_{y_{0}}$ are constants of integration.
It can be checked easily that for $q=0$ our geodesics agree with those of
the region II of the BS metric [8]. It is clear to see that if either $P_{x_{0}}$ or $P_{y_{0}}$
is nonzero then $v$ becomes positive for $ u<\pi/2a$, and particles can pass
from region II to the region IV. Geodesics that remain in region II are
\begin{eqnarray}
x&=&x_{0} \nonumber \\
y&=&y_{0} \nonumber \\
v&=&v_{0} + \frac{\pi \epsilon}{8a}\left(1+3 \sqrt{1+q^{2}}\right)
\end{eqnarray}
where $v_{0} <- \frac{\pi \epsilon}{8a}\left(1+3 \sqrt{1+q^{2}}\right)$. The
effect of cross polarization is that more geodesics remains in region II
relative to the parallel polarization case. On physical grounds
 this result could be anticipated
because cross polarization behaves like rotation which creates a pushing
out effect in the non - inertial frames.
\subsection{Test em field:}
To test the stability of quasiregular singularity, let us consider a test
em field whose vector potential is choosen appropriately as in [9] to be
\begin{equation}
A_{\mu}(v)=\left(0,0,f_{1}(v),f_{2}(v)\right)
\end{equation}
with arbitrary functions $ f_{1}(v)$ and $f_{2}(v)$. The only nonzero
energy - momentum for this test em field is
\begin{equation}
4\pi T_{vv}=-\frac{1}{F\cos^{2}(au)}\left[\left(F^{2}+q^{2}\sin^{2}(au)
 \right)f'^2_1+f'^2_2+2q\sin(au)f'_1 f'_2 \right]
\end{equation}
in which a prime denotes derivative with respect to $v$. Both of scalars
$T^{\mu}_{\mu}$ and $T_{\mu \nu}T^{\mu \nu}$ vanish, predicting that QR
 singularities
are not transformed into a SCS. In the PPON frame.
\begin{eqnarray}
e_{(0)\mu}&=&\left(F,1,0,0 \right) \nonumber \\
e_{(1)\mu}&=&\left(-F,1,0,0 \right) \nonumber \\
e_{(2) \mu}&=&\left(0,0,\cos(au) e^{V/2} cosh\frac{W}{2},-\cos(au)
 e^{-V/2} sinh\frac{W}{2} \right) \nonumber \\
e_{(3) \mu}&=&\left(0,0,-\cos(au) e^{V/2} sinh\frac{W}{2},\cos(au)
 e^{-V/2} cosh\frac{W}{2} \right)
\end{eqnarray}
We find that $T_{(mn)}$ are given in terms of $T_{vv}$ by
\[ 4T_{(mn)}= \left( \begin{array}{cc}
 1&-1\\
 -1& 1
 \end{array} \right)T_{vv} \]
for$ (m,n=0,1)$ and $T_{(mn)}=0$, otherwise. The divergence of this quantity
predicts the occurence of NSCS and therefore QR singularity must be unstable.\\
The stability conjecture therefore correctly finds that these QR singularities
are unstable. However, the same stability conjecture does not find correctly
the nature of the singularity. As we have discussed in section II, the interior
of the interaction region has no SCS. The only SCS is on the null boundaries.
Clarke and Hayward have analysed these singular points for a collinear BS spacetime
and found that the singularity nature of surfaces $(u=0,v=\pi/2b)$ and
 $(v=0,u=\pi/2a)$
are QR. This observation can also be used in the cross polarized version of BS
 spacetime
, because the order of diverging terms in $ \Psi_{0}$ and $\Psi_{4}$ are
 the same.
\subsection{Test Scalar Field:}
The QR singularity structure formed in the incoming region of BS problem
remains unchanged in the case of cross polarized version of the same problem.
Let us now consider the stability of these QR singularities by imposing
a test scalar field in region II which is the one of the incoming region
bounded by the QR singularity. The massless scalar field equation is given
by 
\begin{equation}
\partial _{\mu}\left( g^{\mu\nu}\sqrt{g}\phi_{,\nu}\right)=0
\end{equation}
where we consider $x,y$ independent scalar waves so that a particular
 solution
to this equation is obtained as in the ref()
\begin{equation}
\phi(u,v)=g(u)+\sec(au)f(v)
\end{equation}
where $g(u)$ and $f(v)$ are arbitrary functions. The stress energy tensor
is given by
\begin{equation}
T_{\mu\nu}=\frac{1}{4\pi}\left(\phi_{\mu} \phi_{\nu}-\frac{1}{2}g_{\mu\nu}
 \phi_{\alpha}\phi^{\alpha}\right)
\end{equation}
The corresponding non-zero stress-energy tensors for the test scalar wave is
obtained by taking $g(u)=0$ as,
\begin{eqnarray}
T_{uu}&=&\frac{a^{2}\sec^{2}(au)\tan^{2}(au)f^{2}(v)}{4\pi} \nonumber \\
T_{vv}&=&\frac{\sec^{2}(au)f'^{2}(v)}{4\pi} \nonumber \\
T_{xx}&=&\frac{a\tan(au)f'(v)f(v)}{8\pi F^{2}} \nonumber \\
T_{yy}&=&\frac{a\tan(au)\left[F^{2} +q^{2}\sin^{2}(au)\right]f(v)f'(v)}{
 8\pi F^{2}} \nonumber \\
T_{xy}&=&T_{yx}=\frac{aq\sin(au) \tan(au) f(v)f'(v)}{8 \pi F^{2}}
\end{eqnarray}
It is observed that each component diverges as the QR singularity
 $au\rightarrow\pi/2$
is approached.\\
Next we consider the stress energy tensor in a PPON frame. Such frame vectors
are given in equation (11). The stress-energy tensor is
\begin{equation}
T_{(ab)}=e^{\mu}_{(a)}e^{\nu}_{(b)}T_{\mu\nu}
\end{equation}
The nonzero components are;
\begin{eqnarray}
T_{00}&=&T_{11}=\left(\frac{\sec^{2}(au)}{16 \pi}\right)\left[\frac{a^{2}
 \tan^{2}(au)f^2}{F^{2}}+f'^{2}(v)\right] \nonumber  \\
T_{01}&=&T_{10}=\left(\frac{\sec^{2}(au)}{16 \pi}\right)\left[\frac{a^{2}
 \tan^{2}(au)f^2}{F^{2}}-f'^{2}(v)\right] \nonumber  \\
T_{22}&=&T_{33}=\left(\frac{a\sec^{2}(au)\tan(au)f(v)f'(v)}{8\pi
 F^{3}}\right)\left[F^{2} +2q^{2}\sin^{2}(au)\right] \nonumber \\
T_{32}&=&T_{23}=\frac{aq\sec^{3}(au)\sin^{2}(au)f'(v)f(v)\sqrt{F^{2}
 +q^{2}\sin^{2}(au)}}{4\pi F^{3}}
\end{eqnarray}
These components also diverge as the singularity $au\rightarrow \pi/2$
 is approached.

By the conjecture, this indicates that the QR singularity will be transformed
into a curvature singularity. Finally we calculate the scalar $T_{\mu\nu}
 T^{\mu\nu}$.
\begin{eqnarray*}
T_{\mu\nu}T^{\mu\nu}&=&\frac{a^{2}\sec^{4}(au)\tan^{2}(au)f^{2}(v)
 f'^{2}(v)}{64 \pi^{2} F^{6}}\left\{2F^{4}+2q^{4}\sin^{4}(au)\right.  \\
& & \nonumber \\
& &\left.+\left(F^{2} +q^{2}\sin^{2}(au)\right)\left[13q^{2}\sin^2(au)
 +F^{2}+1\right] \right\}
\end{eqnarray*}
which also diverges as $au \rightarrow \pi/2 $. From these analyses we
 conclude that the curvature singularity formed
will be an SCS. \\

Hence, the conjecture predicts that the QR singularities of cross polarized
 version
of BS spacetime are unstable. It is predicted that the QR singularities
 will be converted to
scalar curvature singularities if generic waves are added.
 The similar results have also been obtained by
HK for the BS spacetime. HK was unable to compare the validity of
 the conjecture
by an exact back-reaction solution. In the next section we present
 an explicit example
that represents cross-polarized em field coupled with scalar field.

\section{Testing The Conjecture For a Class of
 Einstein-Maxwell-Scalar (EMS) Solutions.}
In the former sections, we applied HK stability conjecture to test the
 stability of QR singularities in the
incoming region of CPBS spacetime. According to the conjecture these mild
 singularities are unstable. In order to
see the validity of the conjecture we introduce this new solution.\\

Let the metric be;
\begin{equation}
 ds^{2}=2e^{-M}dudv-e^{-U}[(e^{V}dx^{2} + e^{-V}dy^{2})\cosh W
  -2\sinh W\,dxdy]
\end{equation}
The new solution is obtained from the electrovacuum solution.
 The EMS solution is generated
in the following manner. The Lagrangian density of the system is defined by
\begin{eqnarray}
L&=&e^{-U}\left(M_{u}U_{v}+M_{v}U_{u}+U_{u}U_{v}-W_{u}W_{v}-V_{u}V_{v}
 cosh^{2}W-4\phi_{u}\phi_{v}\right) \nonumber \\
& &-2k\left[\left(B_{u}B_{v}e^{V}+A_{u}A_{v}e^{-V}\right)coshW\right.
 \nonumber \\
& &\left. \hspace{2.5cm}  + \left(A_{u}B_{v}+A_{v}B_{u}\right)sinhW\right]
\end{eqnarray}
which correctly generate all EMS field equations from a variational principle.
 Here
$\phi$ is the scalar field and we define the em potential one-form
 (with coupling constant $k$) by
\begin{equation}
\tilde{A}=\tilde{A_{\mu}}dx^{\mu}=Adx+Bdy
\end{equation}
where $A$ and $B$ are the components in the Killing directions. Variation with
respect to $U,V,M,W,A,B$ and $\phi$ yields the following EMS equations.
\begin{eqnarray}
U_{uv}&=&U_{u}U_{v} \\
2M_{uv}&=&-U_{u}U_{v}+W_{u}W_{v}+V_{u}V_{v}cosh^{2}W+4\phi_{u}\phi_{v} \\
2V_{uv}&=&U_{v}V_{u}+U_{u}V_{v}-2\left(V_{u}W_{v}+V_{v}W_{u}\right)tanhW \nonumber \\
& & -2ksechW\left(\bar{\Phi_{0}}\Phi_{2}+ \bar{\Phi_{2}}\Phi_{0}\right) \\
2W_{uv}&=&U_{v}W_{u}+U_{u}W_{v}+2V_{u}V_{v}coshWsinhW \nonumber \\
& & +2ki\left(\bar{\Phi_{0}}\Phi_{2}- \bar{\Phi_{2}}\Phi_{0}\right) \\
2\phi_{uv}&=&U_{v}\phi_{u}+U_{u}\phi_{v} \\
2A_{uv}&=&V_{v}A_{u}+V_{u}A_{v}-tanhW\left(W_{v}A_{u}+W_{u}A_{v}\right) \nonumber \\
& & -e^{V}\left[2A_{uv}tanhW +W_{u}A_{v}+W_{v}A_{u}\right]  \\
2B_{uv}&=&-V_{v}B_{u}-V_{u}B_{v}-tanhW\left(W_{v}B_{u}+W_{u}B_{v}\right) \nonumber \\
& & -e^{V}\left[2B_{uv}tanhW +W_{u}B_{v}+W_{v}B_{u}\right]
\end{eqnarray}
where $\Phi_{0}$ and $\Phi_{2}$ are the Newmann-Penrose spinors for em plane
 waves
 given as follows
\begin{eqnarray}
\Phi_{2}&=& \frac{e^{U/2}}{\sqrt{2}}\left[e^{-V/2}\left(isinh\frac{W}{2}
 -cosh\frac{W}{2}\right)A_{u}\right. \nonumber \\
& &\left.+e^{V/2}\left(icosh\frac{W}{2}-sinh\frac{W}{2}\right)B_{u}\right] \nonumber \\
\Phi_{0}&=& \frac{e^{U/2}}{\sqrt{2}}\left[e^{-V/2}\left(isinh\frac{W}{2}
 +cosh\frac{W}{2}\right)A_{v}\right. \nonumber \\
& &\left.+e^{V/2}\left(icosh\frac{W}{2}+sinh\frac{W}{2}\right)B_{v}\right] 
\end{eqnarray}
The remaining two equations which do not follow from the variations, namely 
\begin{eqnarray}
2U_{uu}-U^{2}_{u}+2M_{u}U_{u}&=&W^{2}_{u}+V^{2}_{u}cosh^2W+4\phi^{2}_{u}
 +4k\Phi_{2}\bar{\Phi_{2}} \nonumber \\
2U_{vv}-U^{2}_{v}+2M_{v}U_{v}&=&W^{2}_{v}+V^{2}_{v}cosh^2W+4\phi^{2}_{v}
 +4k\Phi_{0}\bar{\Phi_{0}}
\end{eqnarray}
are automatically satisfied by virtue of integrability equations.\\
The metric function $M$ can be shifted now in accordance with \\
\begin{equation}
M=\tilde{M} + \Gamma
\end{equation}
where
\begin{eqnarray}
\Gamma_{u}U_{u}&=&2\phi^{2}_{u} \nonumber \\
\Gamma_{v}U_{v}&=&2\phi^{2}_{v}
\end{eqnarray}
and $U,V,\tilde{M},W$ satisfy the electrovacuum EM equations.
 Integrability condition for the equation (31)
imposes the constraint,
\begin{equation}
\left(\phi_{u}U_{v}-\phi_{v}U_{u}\right)\left[2\phi_{uv}-U_{v}\phi_{u}
 -U_{u}\phi_{v}\right]=0
\end{equation}
from which we can generate a large class of EMS solution.
As an example, for any $ \phi$ satisfying the massless scalar field equation
 the
corresponding $\Gamma$ is obtained from
\begin{equation}
\Gamma=2\int\frac{\phi^{2}_{u}}{U_{u}}\,du+2\int\frac{\phi^{2}_{v}}{U_{v}}\,
 dv
\end{equation}
The only effect of coupling a scalar field to the cross polarized em wave is to
alter the metric into the form,
\begin{equation}
ds^{2}=Fe^{-\Gamma}\left(\frac{d\tau^{2}}{\Delta}-\frac{d\sigma^{2}}{\delta}
 \right)-\Delta F dy^{2}-\frac{\delta}{F}\left(dx-q\tau dy \right)^{2}
\end{equation}
Here $F,\Delta, \delta$ and $q$ represents the solution of electrovacuum
 equations and
$\Gamma$ is the function that derives from the presence of the scalar field.\\

It can be easily seen that for $q=0$ our solution represents pure em
BS solution coupled with scalar field. It constitutes therefore an exact back
 reaction solution to the test
scalar field solution in the BS spacetime considered by HK (). It is clear
 to see that the Weyl scalars
are nonzero and SCS is forming on the surface $ au+bv=\pi /2$. This is in
aggrement with the requirement of stability conjecture introduced by HK. For
$q\neq 0$ the obtained solution forms the exact back reaction solution of the test scalar
field solution in the CPBS spacetime. In Appendix B, we present the Weyl and
 Maxwell scalars explicitly.\\
From the explicit solutions we note that, the Coulomb component $\Psi_{2}$
 remains regular but
$\Psi_{4}$ and $ \Psi_{0}$ are singular when $\tau=1$ or $\sigma=1$. This
 indicates that the singularity structure of the new solution
is a typical SCS. This result is in complete agreement with the stability
 conjecture.

\section{Oppositely Moving Null Dusts In CPBS Spacetime}
{\bf A)} Let us assume first null test dusts moving in the CPBS background.
 For
simplicity we consider two different cases the $x=\mbox{constant}$ and
 $y=\mbox{constant}$ projections
of the spacetime. We have in the first case
\begin{equation}
ds^{2}=\frac{e^{-M}}{2ab}\left(dt^{2}-dz^{2}\right)-e^{-U-V}coshW \,dy^{2}
\end{equation}
where we have used the coordinates $(t,z)$ according to
\begin{eqnarray}
t&=&au+bv \nonumber \\
z&=&au-bv
\end{eqnarray}
The  energy-momentum tensor for two oppositely moving null dusts can
be chosen as
\begin{equation}
T_{\mu \nu}=\rho_{l}l_{\mu}l_{\nu}+\rho_{n}n_{\mu}n_{\nu}
\end{equation}
where $\rho_{l}$ and $\rho_{n}$ are the finite energy densities of the dusts.
 The null propagation directions $l_{\mu}$ and $n_{\nu}$ are
\begin{eqnarray*}
l_{\mu}&=&\left(a_{0},0,a_{2},a_{3}\right) \\
n_{\mu}&=&\left(-a_{0},0,a_{2},a_{3}\right)
\end{eqnarray*}
with
\[
a_{2}=k_{2}=\mbox{constant} \hspace{3cm} a_{3}=\frac{k_{1}}{2ab}=
 \mbox{constant}
\]
\[a_{0}=\frac{1}{2ab}\left(k^{2}_{1}+\frac{2abk^{2}_{2}}{coshW}\,
 e^{U+V-M}\right)^{1/2}\]
We observe from (1) that
\begin{equation}
\frac{e^{U+V}}{coshW} = \frac{F}{\Delta F^{2} + \delta q^{2} \tau^{2}}
\end{equation}
 which is finite for $q \neq 0$. The components of energy-momentum tensors
  in PPON
 frames are proportional to the expression (38). This proportionality makes
  all the
 components of energy-momentum tensors finite.
  In the limit as $q \rightarrow 0 $ which reduces
our line element to the BS this expression diverges on the horizon
 $(\tau = 1)$
. Trace of the energy-momentum is obviously zero therefore we must extract
 information from the scalar
 $ T_{\mu \nu}T^{\mu\nu}$. One obtains,
\begin{equation}
T_{\mu \nu}T^{\mu \nu}=2\rho_{l} \rho_{n} \left(\frac{k^{2}_{1} e^{M}}{ab}
 +\frac{2k^{2}_{2}e^{U+V}}{coshW}\right)^{2}=\mbox{finite}
\end{equation}
The projection on $y=\mbox{constant}$, however is not as promising as the
 $x=\mbox{constant}$
case. Consider the reduced metric
\begin{equation}
ds^{2}=\frac{e^{-M}}{2ab}\left(dt^{2}-dz^{2}\right)-e^{-U+V}coshW\,dx^{2}
\end{equation}
A similar analysis with the null vectors
\begin{eqnarray*}
l_{\mu}&=&\left(a_{0},a_{1},0,a_{3}\right) \\
n_{\mu}&=&\left(-a_{0},a_{1},0,a_{3}\right)
\end{eqnarray*}
with
\[
a_{1}=k_{3}=\mbox{constant} \hspace{3cm} a_{3}=\frac{k_{1}}{2ab}=
 \mbox{constant}
\]
\[a_{0}=\frac{1}{2ab}\left(k^{2}_{1}+\frac{2abk^{2}_{3}}{coshW}\,
 e^{U-V-M}\right)^{1/2}\]
yields the scalar
\begin{equation}
T_{\mu \nu}T^{\mu \nu}=2\rho_{l} \rho_{n} \left(\frac{k^{2}_{1} e^{M}}{ab}
 +\frac{2k^{2}_{3}e^{U-V}}{coshW}\right)^{2}
\end{equation}
From the metric (1) we see that
\begin{equation}
\frac{e^{U-V}}{coshW} = \frac{F}{\delta}
\end{equation}
which diverges on the horizon $ \sigma=1 $. The scalar $ T_{\mu \nu}
 T^{\mu \nu} $
constructed from the test dusts therefore diverges. The PPON components of
 the energy
momentum tensors are also calculated and it is found that all
of the components are proportional to the expression (42).
This indicates that the components of energy-momentum tensor diverges as the
singularity is approached. When we consider the HK stability conjecture an
 exact
back reaction solution must give a singular solution. We present now an exact
 back reaction
solution of two colliding null shells in the interaction region of the CPBS
 spacetime. \\
{\bf B)} The metric given by
\begin{equation}
ds^{2}=\frac{1}{\phi^{2}}\left(2dudv-dx^{2}-dy^{2}\right)
\end{equation}
where $\phi=1+\alpha u \theta(u) + \beta v \theta (v)$ with
 $(\alpha , \beta )$ positive
constants was considered by Wang [12] to represent collision of two null
 shells (or impulsive dusts). The
interaction region is transformable to the de Sitter cosmological spacetime.
 In
other words the tail of two crossing null shells is energetically equivalent
 to the
de Sitter space. It can be shown also that the choice of the conformal factor
 $\phi=1+\alpha u \theta(u) - \beta v \theta (v)$, with $(\alpha , \beta)$
  positive
 constants becomes isomorphic to the anti - de Sitter space. \\

 The combined metric of CPBS and colliding shells can be represented by
 \begin{equation}
 ds^{2}=\frac{1}{\phi^{2}}\,ds^{2}_{CPBS}
 \end{equation}
This amounts to the substitutions
\begin{eqnarray}
M&=&M_{0}+2\ln \phi \nonumber \\
U&=&U_{0}+2\ln \phi \nonumber \\
V&=&V_{0} \nonumber \\
W&=&W_{0}
\end{eqnarray}
where $ \left(M_{0},U_{0},V_{0},W_{0}\right)$ correspond to the metric
 functions
of the CPBS solution. Under this substitutions the scale invariant Weyl
 scalars remain invariant
( or at most multiplied by a conformal factor ) because $ M-U=M_{0}-U_{0}$
 is the
combination that arise in those scalars. The scalar curvature, however,
 which was zero
in the case of CPBS now arises as nonzero and becomes divergent as we
 approach
the horizon. Appendix C gives the scalar quantities $ \Lambda , \Phi_{11}
 ,\Phi_{00} ,\Phi_{22} $ and
$\Phi_{02}$. Thus the exact back reaction solution is a singular one
 as predicted
by the conjecture. It is further seen that choosing $ \beta = 0$,
  which removes one of the
shells leaves us with a single shell propagating in the $v$- direction.
 From the scalars we
see that even a single shell gives rise to a divergent back reaction by
 the spacetime. The
horizon, in effect, is unstable and transforms into a singularity in the
 presence of
two colliding, or even a single propagating null shell. Let us note as an
 alternative
interpretation that the metric (43) may be considered as a colliding em
 waves in a de Sitter background.
Collision of em waves creates an unstable horizon in the de Sitter space
 which is otherwise regular
for $ u>0$ and $v>0$.

\section{Discussion}
In this paper we have analysed the stability of QR singularities in the
 CPBS spacetime.
Three types of test fields are used to probe the stability. First we have
 applied test em field to the
background CPBS spacetime. From the analyses we observed that the QR
 singularity in the incoming
region becomes unstable according to the conjecture, and it is
 transformed into
NSC singularity. This is the prediction of the conjecture. However,
the exact back-reaction solution shoes that beside the true singularities
on the null boundaries  $ u=0, v=\frac{\pi}{2b}$ and $ v=0, u=\frac{\pi}{2a}$.
 There
are quasiregular singularities in the incoming regions. The interior
of interaction region is singularity free and the hypersurface
 $ au +bv=\pi/2$ is a
Killing-Cauchy horizon. As it was pointed out by HK in the case of colliding
em step waves, conjecture fails to predict the correct nature of the
 singularity in the interaction region. We
have also discovered the same behaviour for the cross polarized version of
 colliding em step waves. The addition of cross polarization
 does not alter the existing property. \\

Secondly we have applied test scalar field to the background CPBS spacetime.
 The effect of scalar field on the QR singularity is stronger than the effect
  of em test
field case. We have obtained that the QR singularity is unstable and transformes
 into a SCS. To check the validity of the conjecture , we have constructed a
  new solution constituting
an exact back reaction solution to the test scalar field in the CPBS
 spacetime. The solution
represents the collision of cross polarized em field coupled to a scalar
 field. An examination
of Weyl and Maxwell scalars shows that $ \Psi_{0} , \Psi_{4} , \Phi_{00}
 $ and $ \Phi_{22}$ diverge as the singularity
is approached and unlike the test em field case the conjecture predicts
 the nature
of singularity formed correctly. \\

Finally, we have introduced test null dusts into the interaction region
 of CPBS spacetime.
The conjecture predicts that the CH are unstable and transforms into SCS.
 This
result is compared with the exact back-reaction solution and observed that
 the conjecture works.
\newpage

\section*{Appendix A: \\ Riemann Components for Region II }
To determine the type of singularity in the incoming region of CPBS spacetime,
 the Riemann tensor
 in local and in PPON frame must be evaluated. Non-zero Riemann tensors in
  local
 coordinates are found as follows.
 \begin{eqnarray}
 -R_{uxux}&=& e^{V-U}\left[ \Phi_{22} coshW + (Im \Psi_4 ) sinhW + Re \Psi_4
  \right] \nonumber \\
 -R_{uyuy}&=& e^{-U-V} \left[ \Phi_{22} coshW + (Im \Psi_4 ) sinhW - Re \Psi_4
  \right] \nonumber \\
R_{uxuy}&=& e^{-U}\left[ \Phi_{22} sinhW + (Im\Psi_4)coshW\right]
\end{eqnarray}
where
\begin{eqnarray*}
Re\Psi_4&=&-\frac{1}{2}\left[ \left(V_{uu}-V_uU_u + M_uV_u \right) coshW
 +2V_uW_usinhW\right] \nonumber \\
Im \Psi_4&=&-\frac{i}{2}\left(W_{uu} -U_uW_u +M_uW_u -V^2_u coshW sinhW
 \right) \nonumber \\
\Phi_{22}&=&\frac{1}{4}\left(2U_{uu}-U^2_u - W^2_u - V^2_u cosh^2W \right)
\end{eqnarray*}
Note that in region II the Weyl scalar $ \Psi_4=0 $, therefore only
 $ \Phi_{22}$ exists.
It is clear that $ e^{-U}=0 $ in the limit $ au \rightarrow \pi/2 $, so that
  all of the components vanish
\[ R_{uxuy}=R_{uyuy}=R_{uxux}=0 \]
To find the Riemann tensor in a PPON frame, we define the following PPON
 frame vectors;
\begin{eqnarray*}
e^\mu_{(0)}&=&\left(\frac{1}{2F},\frac{1}{2},0,0\right) \\
e^\mu_{(1)}&=&\left(\frac{1}{2F},-\frac{1}{2},0,0\right) \\
e^\mu_{(2)}&=&\left(0,0,-e^{\frac{U-V}{2}}cosh\frac{W}{2},
 -e^{\frac{U+V}{2}}sinh\frac{W}{2}\right) \\
e^\mu_{(3)}&=&\left(0,0,-e^{\frac{U-V}{2}}sinh\frac{W}{2},
 -e^{\frac{U+V}{2}}cosh\frac{W}{2}\right)
\end{eqnarray*}
In this frame the non-zero components of the Riemann tensors are:
\begin{eqnarray}
R_{0202}&=&-\frac{1}{4F^2}\left(\Phi_{22} + Re\Psi_4\right) \nonumber \\
R_{0303}&=&-\frac{1}{4F^2}\left(\Phi_{22} - Re\Psi_4\right) \nonumber \\
R_{1313}&=&-\frac{1}{4F^2}\left(\Phi_{22} - Re\Psi_4\right) \nonumber \\
R_{1212}&=&-\frac{1}{4F^2}\left(\Phi_{22} + Re\Psi_4\right) \nonumber \\
R_{0302}&=&R_{1312}=\frac{1}{4F^2} Im \Psi_4
\end{eqnarray}
In the limit of $ au \rightarrow \pi/2 $, we have the following results
\begin{eqnarray*}
R_{0202}&=&R_{1212}=R_{0303}=R_{1313}=-\frac{a^2}{(1+q^2)^{3/2}} \\
R_{0302}&=&R_{1312}=0
\end{eqnarray*}
which are all finite. This indicates that the apparent singularity in region
 II (one of the incoming region) is a
quasiregular singularity.
\section*{Appendix B: \\ Properties of the EMS Solution  }
In order to calculate the Weyl and Maxwell scalars we make use of the
CX line element
\begin{equation}
ds^2=U^2\left(d\psi^2 -d\theta^2\right)-\frac{\sin\phi \sin\theta}{1-
 \epsilon\overline{\epsilon}}\, \vline (1-\epsilon)dx+i(1+\epsilon)dy \,
  \vline^{2}
\end{equation}
where
\begin{eqnarray*}
U^2&=&F e^{-\Gamma} \hspace{2.5cm} \epsilon=\frac{Z-1}{Z+1} \\
Z&=&\sin\theta \left(\Sigma \sin\psi - i\sin \alpha \sin \theta \cos \psi
 \right)^{-1}
\end{eqnarray*}
$\Sigma $ is given in equation (4) and we have chosen $ a=1=b$, such that
the new coordinates $(\theta, \psi)$ are related to $ (\tau, \sigma)$
by
\begin{eqnarray*}
\tau&=&\cos \psi \\
\sigma&=&\cos \theta
\end{eqnarray*}
The Weyl and Maxwell scalars are found as
\begin{eqnarray}
\Psi_2&=&Re^{\Gamma} \\
\Psi_4&=&-e^{\Gamma + i\lambda} \left[ 3R
 +\frac{1}{4F\Sigma \sin \theta \sin \psi}
  \left(\Sigma \sin(\theta+\psi)
  -\Sigma_{\theta} \sin \psi \sin\theta \right. \right. \nonumber \\
& &   \nonumber \\
& &\left.\left.+ i\sin \alpha \sin^2 \theta \sin \psi \right)
\left(\Gamma_{\psi}-\Gamma_{\theta}\right)\right] \\
& & \nonumber \\
\Psi_0&=&-e^{\Gamma - i\lambda} \left[ 3R
 +\frac{1}{4F\Sigma \sin \theta \sin \psi} \left(\Sigma \sin(\psi
-\theta)-\Sigma_{\theta} \sin \psi \sin\theta \right. \right. \nonumber \\
& & \nonumber \\
& &\left.\left.+ i\sin \alpha \sin^2 \theta \sin \psi \right)
\left(\Gamma_{\psi}+\Gamma_{\theta}\right)\right] \\
& & \nonumber \\
2\Phi_{00}&=& e^{\Gamma}\left[\frac{\cos \alpha}{\Sigma^2}-\frac{
 \sin(\psi+\theta)(\Gamma_{\theta}+\Gamma_{\psi})}{2F\sin \psi \sin \theta}
  \right] \\
& & \nonumber \\
2\Phi_{22}&=& e^{\Gamma}\left[\frac{\cos \alpha}{\Sigma^2}-\frac{\sin(\theta-
 \psi)(\Gamma_{\theta}-\Gamma_{\psi})}{2F\sin \psi \sin \theta} \right] \\
& & \nonumber \\
-2\Phi_{02}&=&e^{\Gamma + i\lambda}\frac{\cos \alpha}{\Sigma^2}
\end{eqnarray}
where
\begin{eqnarray*}
R&=&\left(\frac{\sin(\alpha/2)}{2\Sigma}\right)\left[\frac{\sin(\alpha/2)
-i\cos \theta \cos(\alpha/2)}{\cos(\alpha/2) + i\cos \theta \sin(\alpha/2)}
 \right] \\
& & \\ 
e^{i\lambda}&=&\frac{\sin \theta + \Sigma \sin \psi + i \sin \psi \sin \theta
 \cos \psi}
{\sin \theta + \Sigma \sin \psi - i \sin \psi \sin \theta \cos \psi}
\end{eqnarray*}
\section*{Appendix: C \\ The Weyl and Maxwell Scalars }
The non-zero Weyl and Maxwell scalars for the collision of null shells in
the background of CPBS spacetime are found as follows.
\begin{eqnarray}
4\phi e^{-M}\Phi_{11}&=& \left[ (a\beta + \alpha b)\tan(au+bv) \right.
 \nonumber \\
& & \nonumber \\
& & \left.+(a \beta-\alpha b)\tan(au-bv) \right] \theta(u) \theta(v) \\
& & \nonumber \\
4\phi e^{-M}\Lambda&=& \left[ (a\beta + \alpha b)\tan(au+bv)+(a \beta
-\alpha b)\tan(au-bv)\right. \nonumber \\
& & \nonumber \\
& & \left.+\frac{4\alpha \beta}{\phi} \right] \theta(u) \theta(v) \\
& & \nonumber \\
\Phi_{22}&=&(\Phi_{22})_{CPBS} + \left(\frac{\alpha e^M }{\phi}\right)\left[
 \delta(u) \right.\nonumber \\
& & \nonumber \\
& &\left. - \theta(u)\left(a \Pi +\frac{u}{(1-u^2)(1-v^2)}\right) \right]\\
& & \nonumber  \\
\Phi_{00}&=&(\Phi_{00})_{CPBS} + \left(\frac{\beta e^M }{\phi}\right)\left[
 \delta(v) \right. \nonumber  \\
& & \nonumber \\
& & \left.+ \theta(v)\left(b \Pi - \frac{v}{(1-u^2)(1-v^2)}\right)\right] \\
& & \nonumber \\
\Phi_{02}&=& (\Phi_{02})_{CPBS}+\left( \frac{e^M}{4FY\phi}\right)
\left[\frac{1}{F}\left(\alpha Q \theta(u) + \beta P \theta(v)\right) \right.
 \nonumber \\
& & \nonumber \\
& & \left. +iq\left(\alpha L \theta(u) + \beta K \theta(v)\right)\right]
\end{eqnarray}
where
\begin{eqnarray*}
\phi&=&1+\alpha u \theta(u) + \beta v \theta (v) \\
Q&=&b\left[2q^2\sin(au+bv)\cos(au-bv)-F^2\left(\tan(au-bv)+\tan(au+bv)\right)
 \right. \\
& & \\
& & \left. -2F\cos(au-bv)\sin(au-bv)\left(\sqrt{1+q^2}-1\right)\right] \\
& & \\
P&=&a\left[2q^2\sin(au+bv)\cos(au-bv)+F^2\left(\tan(au-bv)-\tan(au+bv)\right)
 \right. \\
& & \\
& & \left. +2F\cos(au-bv)\sin(au-bv)\left(\sqrt{1+q^2}-1\right)\right] \\
& & \\
Y&=&\left(1+\frac{q^2}{F^2}\tan(au+bv)\sin(au+bv)\cos(au-bv)\right)^{1/2}\\
& & \\
K&=&\frac{a}{\sqrt{\cos(au+bv)\cos(au-bv)}}\left[\frac{\cos(au-bv)}{
 \cos(au+bv)}+\sin2au \right. \\
& & \\
& & \left. -\frac{2\left(\sqrt{1+q^2}-1\right)\sin(au+bv)\cos(au-bv)
 \tan(au-bv)}{F}\right] \\
& & \\
L&=&\frac{b}{\sqrt{\cos(au+bv)\cos(au-bv)}}\left[\frac{\cos(au-bv)}{
 \cos(au+bv)}+\sin2bv \right. \\
& & \\
& & \left. +\frac{2\left(\sqrt{1+q^2}-1\right)\sin(au+bv)\cos(au-bv)
 \tan(au-bv)}{F}\right] \\
& & \\
\Pi&=&\frac{\left(\sqrt{1+q^2}-1\right)\sin(2au-2bv)}{\sqrt{1+q^2}+1+
 \left(\sqrt{1+q^2}-1\right)\sin^2(au-bv)}
\end{eqnarray*}

\newpage
\begin{center}
\bf{\Large Figure caption}
\end{center}

Fig.1(a): Single plane waves with added colliding test fields indicated
 by arrows.
CH exists on the surface $u=1$. \\
(b): Colliding plane wave spacetime with CH's in the incoming regions at
 $ u=1 $ and
$ v=1 $. Test fields are added to test the stability of CH existing in
 region IV.

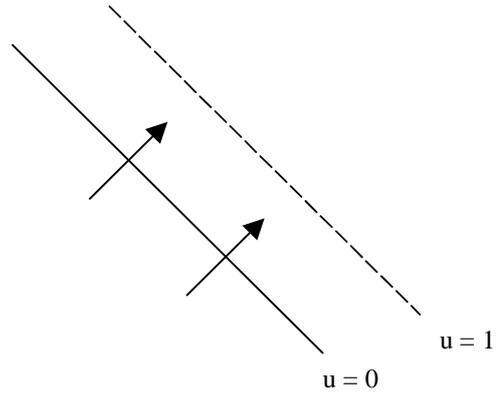
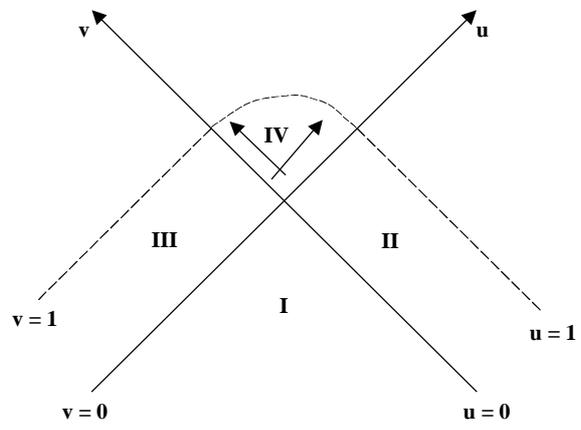
\end{document}